 \documentclass[preprint,authoryear,5p,times,twocolumn]{elsarticle}

\usepackage{graphics}
\usepackage{epsfig}

\usepackage{amssymb}
\usepackage{amsthm}

\journal{New Astronomy}

\begin{document}

\begin{frontmatter}



\title{Electromagnetic counterparts from counter-rotating
  relativistic kicked discs}


\author[oz]{O. Zanotti\corref{cor1}}

\address[oz]{Max-Planck-Institut f{\"u}r
  Gravitationsphysik, Albert Einstein Institut, Am
  M{\"u}hlenberg 1, 14476 Golm, Germany}

\cortext[cor1]{Corresponding author}
\ead{zanotti@aei.mpg.de}

\begin{abstract}
We show the results of two dimensional general
relativistic inviscid and isothermal hydrodynamical simulations comparing the
behavior of co-rotating (with respect to the black hole
rotation) and counter-rotating circumbinary
quasi-Keplerian discs in the
post merger phase of a supermassive binary black hole
system. While confirming the spiral shock generation
within the disc due to the combined effects of mass loss
and recoil velocity of the black hole, we find that the
maximum luminosity of counter-rotating discs is 
a factor $\sim(2-12)$ 
higher than in
the co-rotating case, depending on the spin of the
black hole. On the other hand, 
the luminosity peak happens 
$\sim 10$ days later with respect to the co-rotating
case.
Although the global dynamics of
counter-rotating discs in the post merger phase of a
merging event is very similar to that for
co-rotating discs, an important difference has been
found. 
In fact,
increasing the spin of the
central black hole produces more luminous co-rotating
discs while less luminous counter-rotating ones.

\end{abstract}


\end{frontmatter}


\newcommand{\be}{\begin{equation}}
\newcommand{\ee}{\end{equation}}
\newcommand{\bdm}{\begin{displaymath}}
\newcommand{\edm}{\end{displaymath}}
\newcommand{\bea}{\begin{eqnarray}}
\newcommand{\eea}{\end{eqnarray}}
\newcommand{\PNM}{P_NP_M}
\newcommand{\halb}{\frac{1}{2}}
\newcommand{\FQi}{\tens{\mathbf{F}}\left(\Qi\right)}
\newcommand{\FQj}{\tens{\mathbf{F}}\left(\Qj\right)}
\newcommand{\FQjj}{\tens{\mathbf{F}}\left(\Qjj\right)}
\newcommand{\nj}{\vec n_j}
\newcommand{\FORCE}{\textnormal{FORCE}}
\newcommand{\GFORCE}{\textnormal{GFORCEN}}
\newcommand{\LF}{\textnormal{LF}'}
\newcommand{\LW}{\textnormal{LW}'}
\newcommand{\WL}{\mathcal{W}_h^-}
\newcommand{\WR}{\mathcal{W}_h^+}
\newcommand{\nur}{\boldsymbol{\nu}^\textbf{r} }
\newcommand{\nuf}{\boldsymbol{\nu}^{\boldsymbol{\phi}} }
\newcommand{\nut}{\boldsymbol{\nu}^{\boldsymbol{\theta}} }
\newcommand{\ar}{\phi_1\rho_1}
\newcommand{\arr}{\phi_2\rho_2}
\newcommand{\ur}{u_1^r}
\newcommand{\uf}{u_1^{\phi}}
\newcommand{\ut}{u_1^{\theta}}
\newcommand{\urr}{u_2^r}
\newcommand{\uff}{u_2^{\phi}}
\newcommand{\utt}{u_2^{\theta}}
\newcommand{\ub}{\textbf{u}_\textbf{1}}
\newcommand{\ubb}{\textbf{u}_\textbf{2}}
\newcommand{\RoeMat}{{\tilde A}_{\Path}^G} 
\def\astrobj#1{#1}.




\section{Introduction}
\label{introduction}
In the last few years a lot of attention has been
given to the possibility of detecting the
electromagnetic (EM) counterpart of the 
gravitational wave signal produced 
in the merger event
between two supermassive binary black holes (SMBBHs),
whose  gravitational waves
should be detected by the planned Laser Interferometric
Space Antenna
(LISA)~\citep[see, among the
others,][]{Bogdanovic2011,Haiman2009b,Palenzuela:2009hx,Sesana2009,Stavridis2009,Babak2011}.  

Our current understanding of the physics of these systems
distinguishes between a pre-merger phase
and a post-merger phase. 
The EM emission properties of the pre-merger phase have
been studied by several authors, including
~\citet{Moesta:2009,Bode:2009mt,Chang2009,Shapiro:2009uy,
Liu:2010}.  
\citet{Chang2009}, in particular, showed
that, approximately one day before the merger is
completed, there is a late time precursor brightening from tidal and
viscous dissipation in the inner disc. 
On the other hand, the typical behavior of the
SMBBH system in the post-merger phase is
characterized by two distinct but related physical
effects having to do with the emission of gravitational waves, namely 
an abrupt mass loss and a sudden recoil
of the resultant black hole.
Observationally, the ``recoil event'' has recently found indirect
confirmation through spectropolarimetric
measurements of the quasar \astrobj{E1821+643}~\citep{Robinson2010},
that are consistent with a scattering model
in which the broad-line region has two components, moving
with different bulk velocities away from the
observer.  
The kick velocity and the mass loss of the black
hole are extremely relevant both because they represent an
imprint of the gravitational wave emission, and
because they have
strong dynamical effects on the circumbinary
disc, whose inner radius is  
that possessed at the time when
the binary in-spiral time scale  became shorter than the viscous time scale in
the disc.

When investigating the response of the disc to the mass
loss of the black hole and its recoil velocity, 
an additional physical effect that needs to be taken into
account is that produced by counter-rotation.
The term co-rotation (counter-rotation) has two different meanings when
applied to the
pre-merger phase, when it has to be 
intended as rotation in the same (opposite) sense of the merging
binary, and when applied to the post-merger phase, 
when it has to be 
intended as rotation in the
same (opposite) sense of the spinning resultant black hole.
It is worth recalling that,
on a broader sense, counter-rotating discs are
a relatively common phenomenon in galaxies.
\citet{Coccato2011} report a 
list of galaxies, including the well studied  case of
\astrobj{NGC4550}~\citep{Rubin1992}, where two
co-spatial stellar discs, one orbiting prograde, the other
one orbiting retrograde, have been observed.
In general, large scale counter-rotating stellar discs 
manifest preferentially in early-type spirals,
while $20\%$ of the gas discs in S0
galaxies is thought to counter-rotate.
As far as SMBBHs systems are concerned,
\citet{MacFadyen2008}
speculate about the important dynamical differences that
may be present in a counter-rotating disc during the
pre-merger phase.
Quite recently \cite{Nixon2010} have performed a detailed
analysis, both analytical and numerical, of the retrograde accretion in merging
supermassive black holes, reaching the conclusion 
that the binary orbit, by absorbing negative angular momentum
from the circumbinary disc via gas capture, may become eccentric.
In particular, the binary coalesces once it has absorbed the
angular momentum of a gas mass comparable to that of the
secondary black hole. 

All of these investigations, however, refer to the
pre-merger phase. During the post-merger phase, it is
not clear whether a retrograde disc may have significant effects
on both the dynamics and the EM signal emitted by the system.
Several time dependent numerical investigations of
prograde  circumbinary discs in the post-merger phase
have been performed, both in a
Newtonian~\citep{Lippai:2008,Oneill2009,Corrales2009,Rossi2010} 
and in a relativistic
framework~\citep{Megevand2009,Anderson2009,Zanotti2010}. With
different emphasis and caveats, the majority of these
works have confirmed the possibility of detecting a
significant EM counterpart from SMBBHs.
By extending the analysis of \cite{Zanotti2010}, in this
paper the role of retrograde rotation 
in the post-merger phase is studied through numerical
simulations, 
with the special aim
of comparing the emitted luminosity with respect to the prograde
rotation case.

The paper is organized as follows. In Section~\ref{Initial_conditions}
and in Sec.~\ref{Numerical_method}
the essential information about 
the physical properties of the initial models, 
and about the numerical code adopted in the simulations
are provided.
Section~\ref{Results} is devoted to the presentations of the
results, while Section~\ref{Conclusions} contains a summary of
the work.  A signature $\{-,+,+,+\}$ for the space-time
metric is assumed and Greek letters (running from $0$ to $3$) for
four-dimensional space-time tensor components are used, while Latin letters
(running from $1$ to $3$) will be employed for three-dimensional
spatial tensor components. Moreover, $c=G=1$ and the
geometric system of units is extended
by setting $m_p/k_B=1$, where $m_p$ is the mass of the
proton, while $k_B$ is the Boltzmann constant. In this way the
temperature is a dimensionless quantity.

\section{Initial conditions}
\label{Initial_conditions}

\begin{table*}
\begin{center}
\caption{Main properties of the initial models. From left to right the
  columns report the name of the model, the black hole spin parameter
  $a$, the power-law index $q$ and the parameter ${\cal
    S}$ 
  of the angular momentum
  distribution, 
  the inner and the outer radius of the disc, $r_{\rm in}$ and
  $r_{\rm out}$, the radius of the maximum rest-mass density $r_{\rm c}$,
  the orbital period at the radius of maximum rest-mass
  density $\tau_{\rm c}$, the uniform temperature $T$ 
  assumed during isothermal evolution. The adiabatic index
  $\gamma=5/3$ and the mass of the black hole is
  $M=1.0\times 10^{6}M_{\odot}$  in each model.
}
\label{tab1}
\vspace{0.5cm}
\begin{tabular}{l|ccccccccc}
\hline
\hline
Model & $J/M^2$     & $ q $ & ${\cal S}$ & $r_{\rm in}$ & $r_{\rm   out}$ & $r_{\rm c}$ & $\tau_{\rm c}$ & $T$  \\
        &             &       
&              &     $(M)$         & $(M)$       & $(M)$
&  $(d)$     & $(K)$ \\

\hline

\hline

\texttt{CoRot\_a.0.1} & $0.1$  & $0.4955$ & $1.033$  & $668$  &
$1998$ & $1082$  & $12.76$ & $2.68\times 10^{6}$\\
\texttt{CoRot\_a.0.5} & $0.5$  & $0.4955$ & $1.033$  & $598$  &
$2138$ & $1036$  & $11.95$ & $2.68\times 10^{6}$\\
\texttt{CoRot\_a.0.9} & $0.9$  & $0.4955$ & $1.033$  & $537$  &
$2273$ & $988$  & $11.13$ & $2.68\times 10^{6}$\\
\texttt{CoRot\_a.0.99} & $0.99$  & $0.4955$ & $1.033$  & $524$  &
$2303$ & $977$  & $10.94$ & $2.68\times 10^{6}$\\
\hline
\texttt{CounterRot\_a.0.1} & $0.1$  & $0.4955$ & $-1.034$ & $534$  &
$3270$ & $1104$ & $13.15$ & $2.68\times 10^{6}$\\
\texttt{CounterRot\_a.0.5} & $0.5$  & $0.4955$ & $-1.034$ & $577$  &
$3144$ & $1149$ & $13.95$ & $2.68\times 10^{6}$\\
\texttt{CounterRot\_a.0.9} & $0.9$  & $0.4955$ & $-1.034$ & $622$  &
$3019$ & $1192$ & $14.75$ & $2.68\times 10^{6}$\\
\texttt{CounterRot\_a.0.99} & $0.99$  & $0.4955$ & $-1.034$ & $633$  &
$2991$ & $1202$ & $14.93$ & $2.68\times 10^{6}$\\

\hline
\texttt{CounterRot\_T1} & $0.9$  & $0.4955$ & $-1.034$ & $622$  &
$3019$ & $1192$ & $14.75$ & $4.47\times 10^{5}$\\
\texttt{CounterRot\_T2} & $0.9$  & $0.4955$ & $-1.034$ & $622$  &
$3019$ & $1192$ & $14.75$ & $8.94\times 10^{5}$\\
\texttt{CounterRot\_T3} & $0.9$  & $0.4955$ & $-1.034$ & $622$  &
$3019$ & $1192$ & $14.75$ & $8.05\times 10^{6}$\\
\texttt{CounterRot\_T4} & $0.9$  & $0.4955$ & $-1.034$ & $622$  &
$3019$ & $1192$ & $14.75$ & $1.61\times 10^{7}$\\

\hline
\hline

\end{tabular}
\begin{flushleft}

\end{flushleft}
\end{center}
\end{table*}

The initial model is given by a stationary and axisymmetric configuration
obtained after solving the relativistic Euler equations
in the fixed background spacetime of a Kerr black hole.
Such solution is due to~\citet{Kozlowski1978} while
a detailed description for computing the corresponding equilibrium
models can be found in~\citet{Daigne04}. 
The specific angular momentum of the resulting
geometrically thick disc 
obeys a power law on the equatorial plane, namely
$\ell (r, \theta = \pi/2) = {\cal S} r^q$,
where ${\cal S}$ is positive or negative, corresponding to
a disc rotation that is prograde or retrograde
with respect to the black hole rotation.  
Table~\ref{tab1} reports the main
parameters of the models
considered, differing for the rotation law, 
for the spin of the central black
hole, and for the temperature of the disc. 
Having chosen the exponent $q$ to be close to $1/2$ in
all of the models,
the rotation law tends to the Keplerian one, and the disc flattens
towards the equatorial plane, thus making its vertical 
structure negligible. 
The equation of state of the initial model is 
that of a polytrope $p=\kappa \rho^\gamma$, with
$\gamma=5/3$.
It should be noted that it is not
possible to build two models, one co-rotating and the
other one counter-rotating, while maintaining {\em all} of the other
physical parameters unmodified. In particular,
a specific counter-rotating model is larger than the
corresponding co-rotating model with the same disc temperature.
Another relevant difference among the two classes of
models, which is clearly visible by inspection of Table~\ref{tab1}, 
is that, while for co-rotating models the radius of the
maximum rest mass density decreases for larger spins of
the central black hole, the opposite happens for
counter-rotating models\footnote{Note that, when changing
  the spin of the black hole, the potential gap between
  the position of the cusp and the position of the inner
  radius is kept constant.}.
This will have important
implications on the dependence of the emitted luminosity
on the spin, as discussed in Sec.~\ref{Results}.

The recoil effect on the black hole resulting from a 
SMBBHs merger
was predicted well
before~\citep{Bekenstein1973,Redmount:1989}
numerical relativistic 
simulations were able to prove its existence and measure
its magnitude. 
Because in this contribution only two dimensional
simulations are performed\footnote{It is worth recalling
  that the expansion of the disc along the vertical
  direction is a very small effect, since vertical tangential
  velocities along the shock front are at least one order of magnitude
  smaller than those across the shock.},
the recoil velocity vector has to lie 
on the orbital plane of the disc, which is consistent
with a binary merger between two black holes having the
same mass and with spins that are equal in modulus and
both perpendicular to the orbital angular
momentum~\citep{Campanelli:2007ew,Rezzolla:2008sd}. 
Since  the recoil velocities in the orbital plane are expected to be
$\lesssim 450\,\rm{km}/\rm{s}$~\citep{Koppitz-etal-2007aa, Herrmann:2007ac,
  Pollney:2007ss:shortal}, the conservative value $V_{\rm
  k}=300\,\rm{km}/\rm{s}$ is adopted.
In order to simulate this effect numerically, at time
$t=0$ a Lorentz boost to the fluid
velocity of the disc is applied, oriented 
along the radial direction with $\phi=0$.
In addition,  $2\%$ of the total mass-energy of the
merged black hole is assumed to be radiated away in
gravitational waves. In practice,
the initial model is
first computed in the gravitational potential of
the full black hole mass, and then evolved in the gravitational
potential of the reduced mass. 

\section{Numerical method}
\label{Numerical_method}

The equations of general relativistic hydrodynamics are
solved in the stationary spacetime of a Kerr black hole,
written in Boyer-Lindquist coordinates,
through the \texttt{ECHO}
code~\citep{DelZanna2007}. \texttt{ECHO} adopts a $3+1$ split of
spacetime in which the space-time metric is decomposed according to
\be
\mathrm{d}s^2 = g_{\mu\nu}dx^\mu dx^\nu=\! -\alpha^2\mathrm{d}t^2+\gamma_{ij}\,
(\mathrm{d}x^i\!+\beta^i\mathrm{d}t)(\mathrm{d}x^j\!+\beta^j\mathrm{d}t),
\label{eq:adm}
\ee 
where $\alpha$ is the lapse function, $\beta^i$ is the shift vector,
and $\gamma_{ij}$ is the spatial metric tensor, with $i$
and $j$ running from $1$ to $3$.
Moreover,
\texttt{ECHO} adopts a conservative formulation of
the  general-relativistic inviscid hydrodynamical
equations, by solving the system
\be
\partial_t\vec{\mathcal{U}} + \partial_i\vec{\mathcal{F}}^i=\vec{\mathcal{S}},
\label{eq:UFS}
\ee
where 
the vector of conservative variables $\vec{\mathcal{U}}$
and the corresponding fluxes $\vec{\mathcal{F}}^i$ in
the $i$ direction are respectively given by
\be
\vec{\mathcal{U}}\equiv\sqrt{\gamma}\left[\begin{array}{c}
D \\ \\ S_j \\ \\U
\end{array}\right],~~~
\vec{\mathcal{F}}^i\equiv\sqrt{\gamma}\left[\begin{array}{c}
\alpha v^i D-\beta^i D \\\\
\alpha W^i_j-\beta^i S_j \\\\
\alpha S^i-\beta^i U
\end{array}\right] ,
\label{eq:fluxes}
\ee
whereas the sources, in any stationary background metric,
can be written as
\be
\vec{\mathcal{S}} \equiv \sqrt{\gamma}\left[\begin{array}{c}
0 \\  \\
\frac{1}{2}\alpha W^{ik}\partial_j\gamma_{ik}+
S_i\partial_j\beta^i-U\partial_j\alpha \\ \\
\frac{1}{2}W^{ik}\beta^j\partial_j\gamma_{ik}+{W_i}^j\partial_j\beta^i
-S^j\partial_j\alpha
\end{array}\right] \ .
\ee
The determinant of the
spatial metric $\sqrt{\gamma}$ and of the global metric
$\sqrt{-g}$ are related by  $\sqrt{\gamma}\equiv \sqrt{-g}/\alpha$.
The conservative
variables $(D,S_j,U)$ that are evolved by the numerical
scheme are related to 
the rest-mass density $\rho$, to the thermal
pressure $p$ and to the fluid velocity $v_i$ by
\bea
&&D   \equiv \rho W ,\\
&&S_i \equiv \rho h W^2 v_i, \\
&&U   \equiv \rho h W^2 - p, 
\label{eq:cons}
\eea
where $W=(1-v^2)^{-1/2}$ is the Lorentz factor of the fluid
with respect to the Eulerian observer associated to the $3+1$
splitting of the spacetime, and
\be
W_{ij} \equiv \rho h W^2 v_i v_j +p \gamma_{ij} \\
\label{eq:W} 
\ee
is the fully spatial projection of the energy-momentum
tensor of the fluid.

\begin{figure*}
\centering
{\includegraphics[angle=0,width=8.2cm,height=7.8cm]{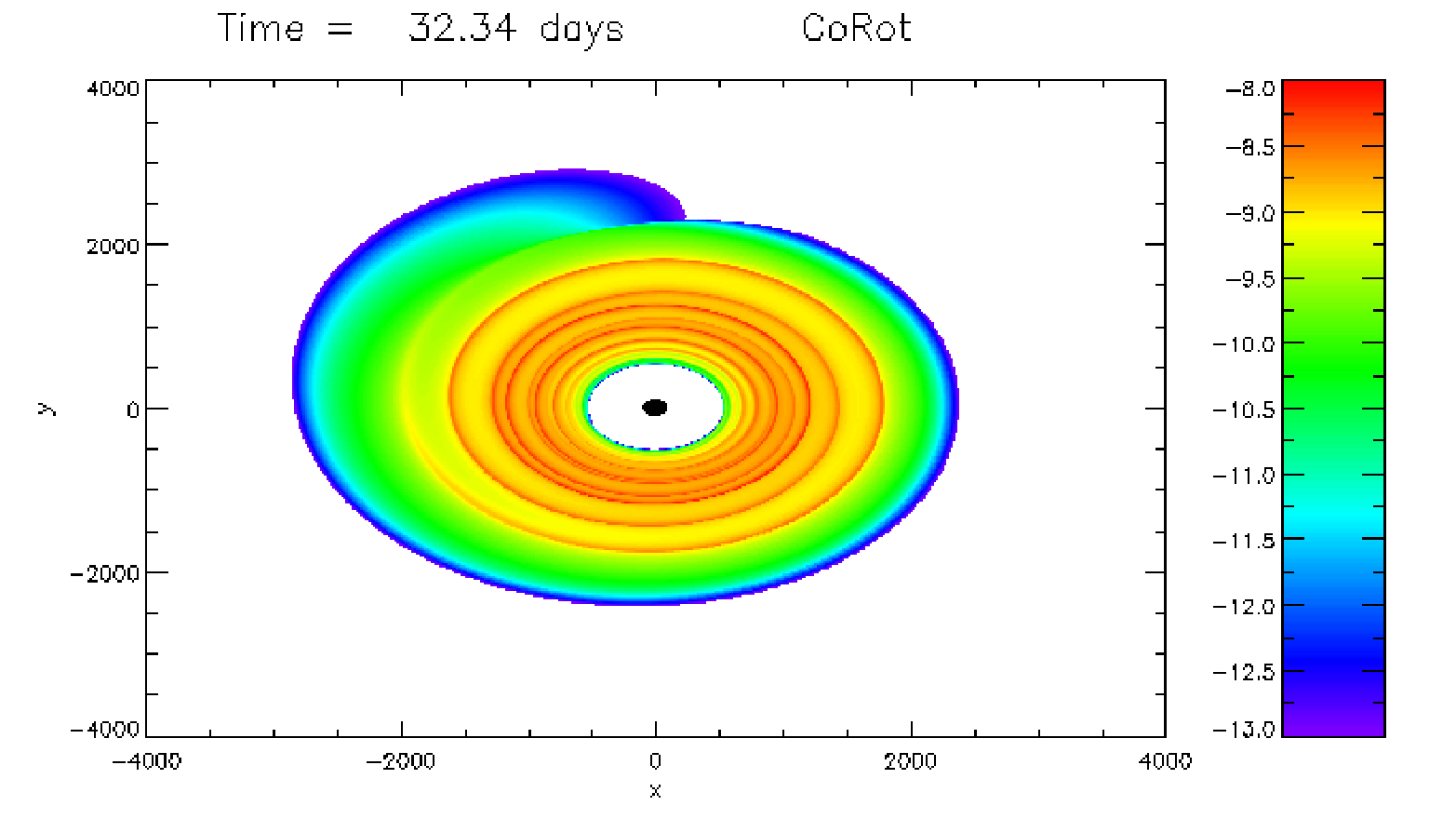}}
{\includegraphics[angle=0,width=8.2cm,height=7.8cm]{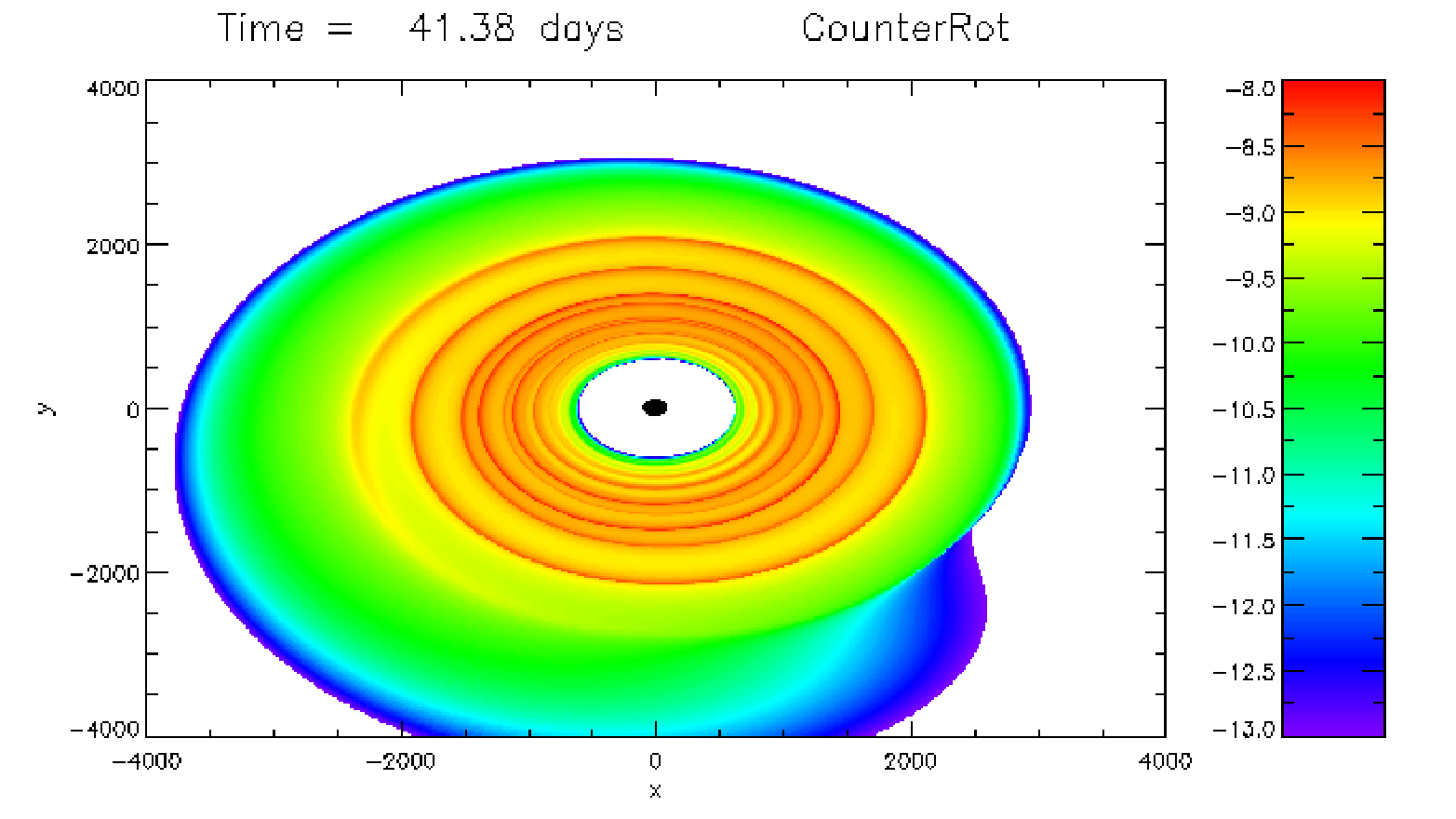}}
\caption{Rest mass density distribution in the
  co-rotating model \texttt{CoRot\_a.0.9} 
(left panel) and in the counter-rotating
  model \texttt{CounterRot\_a.0.9}
(right panel) for a recoil velocity $V_{\rm
    k}=300\,\rm{km}/\rm{s}$ at the time (after merger) when the
  luminosity reaches a maximum. ${\rm cgs}$ units are reported.}
\label{fig1}
\end{figure*}

The formulation above is 
appropriate for numerical integration via standard
high-resolution shock-capturing (HRSC) methods developed for the Euler
equations~\citep{Toro99}.
All the simulations cover a physical
duration much shorter than the viscous time-scale,
making the inviscid approximation a very good one.
In the following, however, we consider an isothermal
evolution of the hydrodynamical equations.
Therefore, the temperature of the disc is assumed to be constant in
space and in time,
and set to the value $T$ reported in Table~\ref{tab1}.
As a result, there is
no need to evolve the energy equation for $U$,
since the energy can be computed directly from the temperature and the
latter is constant by construction. However,
an equation for the time evolution of the
internal energy is actually solved
\bea
\label{internal_energy1}
&&\hskip -1.0cm \partial_t(\sqrt{\gamma}W\rho\epsilon)+\partial_i[\sqrt{\gamma}\rho\epsilon
  W(\alpha
  v^i-\beta^i)] = \nonumber \\ 
&& -p\partial_t(\sqrt{\gamma}W) -p\partial_i(\alpha\sqrt{\gamma}u^i) 
\ ,
\eea
where $\epsilon$ is the
specific internal energy and $p=\rho\epsilon(\gamma-1)$.
The solution of Eq.~(\ref{internal_energy1}) allows to obtain
a good estimate of the emitted luminosity, 
assuming that all the changes in the temperature produced
by local compressions will be dissipated away as radiation. This idea,
proposed in the Newtonian framework by~\citet{Corrales2009}, 
has been first implemented in a general relativistic context
by~\citet{Zanotti2010}.  
The luminosity is then computed in the following way: 
the time-update of Eq.~(\ref{internal_energy1})
provides the value of the quantity $\rho\epsilon$ at time
$t^{n+1}=t^n+\Delta t$, where $\Delta t$ is the time-step
of the simulation. After that, a volume integration
of the difference $\rho[\epsilon - \epsilon(T)]$ is
performed, where $\epsilon(T)=k_BT/[m_p(\gamma-1)]$ is the value prescribed
by the assumption of isothermal evolution.
Finally, division by $\Delta t$ provides the
luminosity.  
The specific internal
energy is reset to $\epsilon = \epsilon(T)$ after each
time-step, so as to guarantee that the
evolution is isothermal. In practice,
Eq.~(\ref{internal_energy1}) is evolved in time
with the only aim of computing the
difference $\rho[\epsilon - \epsilon(T)]$, which is then assumed to be
radiated instantaneously. 

The radial
numerical grid is discretised by choosing $N_r=1200$ points,
non-uniformly distributed from
$r_\mathrm{min}=100M$ to $r_\mathrm{max}=6000M$,
while
outflow boundary conditions are adopted both at
$r_\mathrm{min}$ and $r_\mathrm{max}$.  The azimuthal grid extends
from $0$ to $2\pi$,  with
symmetric (i.e. periodic) boundary conditions adopted at $\phi=0$,
while the number of  angular grid points is $N_\phi=200$. 
All runs are performed with a Courant-Friedrichs-Lewy
coefficient ${\rm CFL}=0.5$.  We have verified through a
series of tests that the choice
$(N_r,N_\phi)=1200\times200$ suffices to find a converged
solution, and, in consequence, all models have been
computed with this canonical resolution.

\section{Results}
\label{Results}

As shown numerically for the first time by \citet{Lippai:2008}, and then 
confirmed repeatedly by several 
authors~\citep{Oneill2009,Corrales2009,Rossi2010,Megevand2009,Anderson2009,Zanotti2010}, 
the combined
effects of mass loss from the central black hole and of the 
recoil velocity generate a spiral shock
pattern that transports angular momentum outwards, and
become responsible of an enhanced
luminosity\footnote{Note, however, that the spiral shock
can form even in the absence of a recoil velocity~\citep{Oneill2009}.}.

Figure~\ref{fig1} shows the iso-density curves of the rest
mass density in the two models \texttt{CoRot\_a.0.9} and
\texttt{CounterRot\_a.0.9},
plotted when the emitted luminosity
reaches its maximum.
Because the recoil velocity is not large, neither of the
two discs has
penetrated into the central cavity at this time. However,
the spiral structure is still clearly visible, producing
regions of high compression. The role of this spiral
pattern has been somewhat overestimated in previous
analysis. Indeed, as shown by~\citet{Zanotti2010} 
through the use a sophisticated shock detector, 
such spiral pattern is not always a true physical
shock. In addition, the same pattern can be obtained even
in the absence of a mass loss or of a recoil velocity,
being produced as the result of non-axisymmetric
hydrodynamical instabilities taking place in the disc.
Therefore, \emph{the spiral pattern cannot be regarded as an
unambiguous imprint of a recoiling black hole.}
It is also worth stressing that,
although the inner edge of the accretion disc is placed
in a region where general relativistic effects may be
regarded as negligible, during the subsequent evolution
(later than the time shown in Fig.~\ref{fig1}),
the central cavity is filled with gas, thus making the
relativistic calculation pertinent.

\begin{figure}
\vspace*{-3.5cm}
{\includegraphics[angle=0,width=8.0cm]{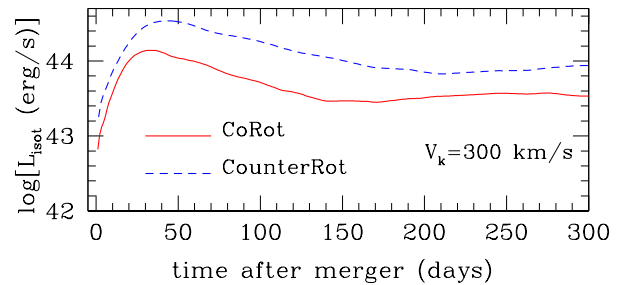}}
\caption{Luminosity computed through the isothermal evolution
  for the co-rotating model \texttt{CoRot\_a.0.9} 
and the counter-rotating one 
\texttt{CounterRot\_a.0.9}.
}
\label{fig2}
\end{figure}

\begin{figure*}
{\includegraphics[angle=0,width=8.0cm]{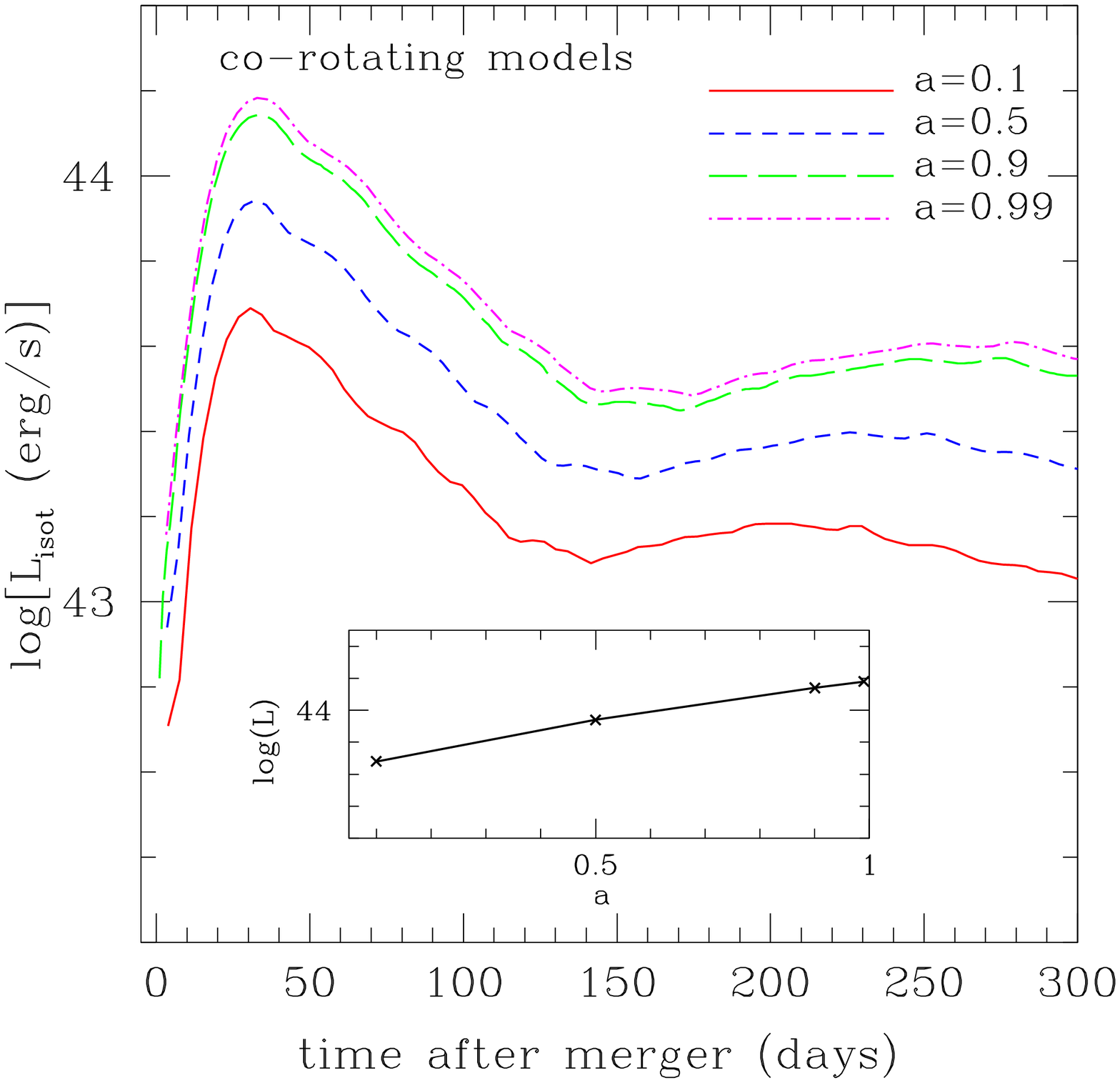}}
{\includegraphics[angle=0,width=8.0cm]{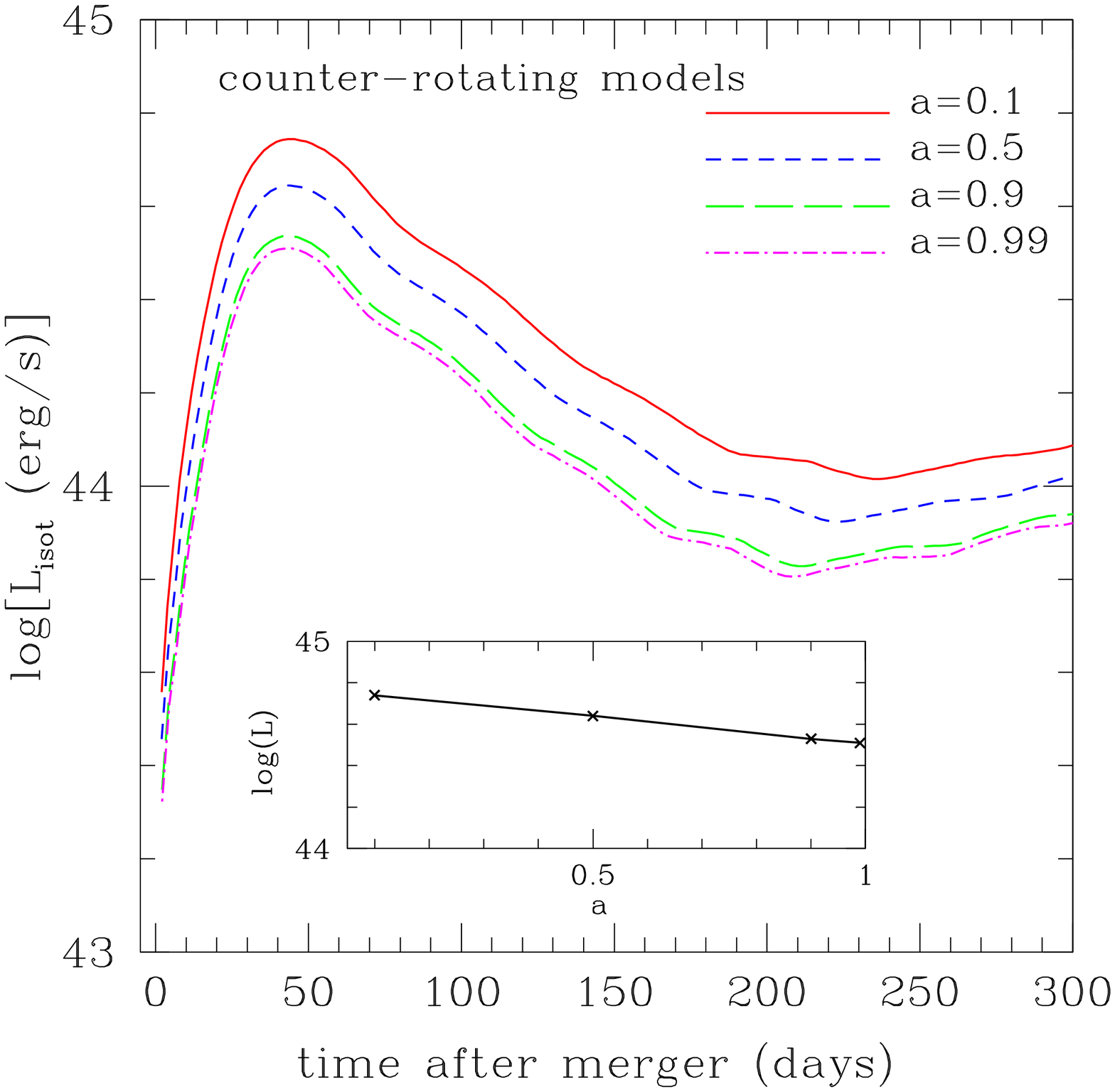}}
\caption{Dependence of the luminosity on the spin of the
  black hole for corotating (left panel) and for 
  counter-rotating (right panel) models. The insets show
  the peak luminosity as a function of the black hole spin.
}
\label{fig3}
\end{figure*}

\begin{figure}
{\includegraphics[angle=0,width=8.0cm]{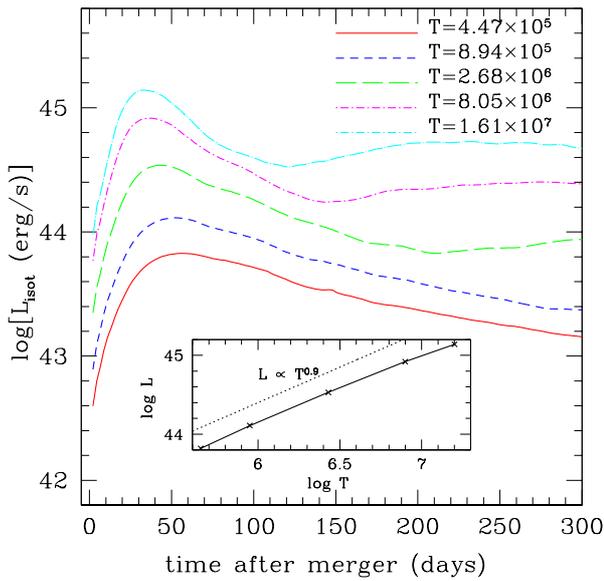}}
\caption{Light curves for different values of the 
disc temperature in counter-rotating models.
}
\label{fig4}
\end{figure}

Figure~\ref{fig2}, on the other hand, shows the
electromagnetic luminosity computed according to the
procedure described in Sec.~\ref{Numerical_method}.
The negative contributions
that are produced in regions experiencing
rarefactions have been neglected.
Because when accounted for these negative contributions
typically yield values that are a factor $\sim 2-5$ smaller, the
values in Fig.~\ref{fig2}-Fig.~\ref{fig4} should be taken as upper limits to the
emitted luminosity~\citep[see][]{Corrales2009}.
More details regarding
this approximation can be found in~\citet{Zanotti2010},
where such an approach is compared to
alternative ones for extracting a luminosity 
in the absence of a fully radiation
hydrodynamics treatment.
The two curves 
refer to the models 
\texttt{CoRot\_a.0.9} and
\texttt{CounterRot\_a.0.9},
having the same
temperature $T=2.68\times10^6 {\rm K}$
and the same polytropic constant, subject to a recoil
velocity $V_{\rm  k}=300\,\rm{km}/\rm{s}$ and with a mass
loss after the merging event that amounts to $2\%$ of
the initial mass of the black hole.
As initially put into evidence by the analysis
of ~\citet{Daigne04}, counter-rotating models have larger
extension with respect to co-rotating ones and, in
addition,  have a larger radius of the maximum rest mass
density point. These two facts
are ultimately responsible for the different behavior
of the light curves reported in Fig.~\ref{fig2}.
Because $r_{\rm c}$ is larger for
\texttt{CounterRot\_a.0.9} than for
\texttt{CoRot\_a.0.9}, the spiral shock that forms after
the merger event takes longer to reach the region of
maximum rest-mass density, which is also the region of
highest compressions where most of the luminosity is emitted. As a result, 
the time of the maximum luminosity is 
$t_{\rm peak}\sim 32$ days for model
\texttt{CoRot\_a.0.9} and 
$t_{\rm peak}\sim 41$ days for model \texttt{CounterRot\_a.0.9}.
Moreover, the maximum luminosity at the peak 
is $L\sim 1.3\times 10^{44} {\rm erg/s}$ for model
\texttt{CoRot\_a.0.9} and 
$L\sim 3.4\times 10^{44} {\rm erg/s}$ for model
\texttt{CounterRot\_a.0.9}. 
In a second and tightly related series of simulations the
dependence of the light curves 
on the spin of the central black hole
has been investigated,
while keeping the
same temperature within the discs. When the spin of the
black hole is increased, the radius of the maximum
rest-mass density $r_{\rm c}$ moves towards the center
for co-rotating models, while it moves towards larger
distances for counter-rotating models (see
Tab.~\ref{tab1}). 
Moreover,  when the spin of the
black hole is increased, the size of the disc increases
for co-rotating models, while it decreases for
counter-rotating models. 
Because of this,
changing the spin of the black hole has opposite effects
on the emitted luminosity
for the two classes of models, as reported in
Fig.~\ref{fig3}. {\em Indeed, the fluid compression, and
therefore the energy released, is larger
if it takes place deeper in the potential well, where the
effects of the spiral shock are stronger.} 
This explains why 
increasing the spin of the
central black hole produces more luminous co-rotating
discs and less luminous counter-rotating ones, as
schematically represented by the insets of
Fig.~\ref{fig3}. 
Tab.~\ref{table:ratio} provides an additional
information, by showing the ratio of the peak luminosity
between counter-rotating and co-rotating discs, in terms
of the spin of the black hole. Because of the
anti-correlation highlighted above, the discrepancy in
the luminosity between the two classes of models is
higher at smaller spins.

\begin{table}
  \caption{First column: black hole spin. Second column: 
   ratio of the peak luminosity 
    between counter-rotating models and co-rotating
    models. Third column: difference (in days) between
    the times of the peak in counter-rotating and in
    co-rotating discs.
 \label{table:ratio}}
\begin{center}
  \begin{tabular}{ccc}
    \hline \hline
   $a$  & $\left(L_{\rm Counter}/L_{\rm
      Corot}\right)_{\rm peak}$ & $\left(t_{\rm
      Counter}-t_{\rm Corot}\right)_{\rm peak}$\\
& & $(d)$ \\
    \hline
$0.1$ & $11.74$ & $11.6$\\
$0.5$ & $5.12$  & $11.5$\\
$0.9$ & $2.45$  & $9.5$\\
$0.99$ & $2.13$ & $9.2$\\
\hline
    \hline
  \end{tabular}
\end{center}
\end{table}

Finally, in a last series of simulations the dependence
of the emitted luminosity on the disc temperature has
been considered. The results of these analysis have been
reported in Fig.~\ref{fig4}, which shows the light curves
for five models having the same rotation law and the same
spin of the central black hole, while different disc
temperatures. An almost linear increase of the peak
luminosity (see inset of Fig.~\ref{fig4}) with the disc
temperature has been found, with $L_{\rm peak}\sim
1.3\times 10^{45}{\rm erg/s}$ for $T\sim1.6\times10^7
K$. 
Such a linear increase is not a surprising result,
given the fact that 
the luminosity is computed from local variations of the
specific internal energy and that the latter scales like
$\epsilon\propto T$. 
Finally, the linear increase of the peak
luminosity with the disc
temperature, which has been reported in
Fig.~\ref{fig4} for counter-rotating models, is not
affected by the rotation law, and the same linear
dependence 
has been found also for co-rotating models.

\section{Conclusions}
\label{Conclusions}

We have performed two dimensional (on the orbital plane) 
relativistic isothermal hydrodynamical simulations
of counter-rotating circumbinary discs that react to the mass loss
and to the recoil velocity of the central black hole
after a merging event of supermassive black holes.
These calculations are very relevant 
for correlating the electromagnetic signal
to the gravitational one, in view of the planned Laser
Interferometric Space Antenna (LISA) mission.
In this respect,
after considering the dynamical response of a disc to a
typical recoil velocity 
$V_{\rm k}=300\,\rm{km}/\rm{s}$ oriented on the orbital plane,
the results show that the maximum 
luminosity of counter-rotating discs is a factor $\sim
2$ higher than in co-rotating discs for high spinning
black holes, while it is a factor $\sim12$ higher for
slowly rotating black holes.
On the other hand, 
the peak of the luminosity typically happens 
$\sim 10$ days later with respect to the co-rotating
case, with
only a weak dependence on the spin of the black hole.
All of these effects are due to the different sizes and
characteristic radii in the two classes of models.
In particular, 
when the spin of the resultant black hole is larger, 
co-rotating discs in the post-merger phase
are more luminous, while 
counter-rotating discs are less luminous.

\section*{Acknowledgments}
I would like to thank an anonymous
referee for very useful comments.
The computations presented in this paper were performed 
on the 
National Supercomputer HLRB-II based on  SGI's Altix 4700
platform installed at Leibniz-Rechenzentrum and on the
IBM/SP6 of CINECA (Italy) through the
``INAF-CINECA'' agreement 2008-2010. 
This work was supported in part by the DFG grant
SFB/Transregio~7.

\bibliographystyle{model1c-num-names}







\end{document}